\documentclass[sigconf,screen,nonacm] {acmart}
\AtBeginDocument{%
  }

\setcopyright{cc}
\setcctype[4.0]{by}

\begin{document}

\title{AI as Friction for Reflection Support in Ideation}

\author{Janin Koch}
\orcid{0000-0001-9207-9550}
\affiliation{%
  \institution{Univ. Lille, CNRS, Inria, Centrale Lille, UMR 9189 CRIStAL
  }
  \city{Lille}
  \country{France}
  }
  \email{janin.koch@inria.fr}

\author{Xiaohan Liao}

\affiliation{%
  \institution{Univ. Lille, Inria, CNRS, Centrale Lille, UMR 9189 CRIStAL}
  \city{Lille}
  \country{France}
  }
  \email{xiaohan.liao@inria.fr}

\author{G\'ery Casiez}
\orcid{0000-0003-1905-815X}

\affiliation{%
  \institution{Univ. Lille, CNRS, Inria, Centrale Lille, UMR 9189 CRIStAL}
  \postcode{F-59000}
  \city{Lille}
  \country{France}}
\email{gery.casiez@univ-lille.fr}


\begin{abstract}
Generative AI tools for creative work tend to be designed around the goal of removing friction, on the assumption that smoother iteration and faster output translate into more value for the designer. 
We argue, however, that this framing leaves out something important about how design ideation works, namely reflection-in-action.
The act of accepting, rejecting and reworking candidate ideas is both a path to a final outcome and the process through which designers develop the rationale that allows them to think with their ideas and to communicate them to others.
This becomes particularly important in group ideation, where ideas need to be expressed and explained to others to allow the group to extend, reject or combine them further. 
We suggest that AI in design ideation might be more usefully thought of as a friction agent for reflection rather than as a smoothing agent for output. 
This reframing opens up a different role for AI in design ideation, one that supports designers in building and carrying rationale rather than substituting for it.
\end{abstract}




\maketitle
\textbf{Reference:}\newline
Janin Koch, Xiaohan Liao, and Géry Casiez. 2026. AI as Friction for Reflection Support in Ideation. In \textit{Proceedings of The First Reflection in Creative Experience (RiCE) Workshop (RiCE W1)}. ACM Creativity \& Cognition 2026, London, UK.

\section{Introduction}

Generative AI tools for creative work tend to be designed around the goal of removing friction~\cite{Choudhury_Eisenbart_Kuys_2025}. 
Faster iteration, lower prompt cost, smoother output and easier disposal of unwanted candidates are commonly treated as overall improvements~\cite{ali2023using}, on the assumption that less friction means more creative throughput and therefore more value for the designer. 
This framing is intuitive, but it overlooks the crucial part of reflection in design ideation that matters in practice.

In design ideation, the act of accepting, rejecting and reworking ideas is not only a path to a final outcome, it is also how designers build the rationale that allows them to think with their ideas and to communicate them to others. 
When AI smooths this process away, it can leave designers with many candidates and rather little account of why any of them should be preferred or their relation to the underlying problem. 
This becomes particularly visible in group ideation, where designers depend on articulable rationale to defend, contest and combine contributions, and where artifacts that arrive without their underlying reasoning struggle to integrate into the group's evolving understanding of the problem.
Building rationale and bringing it into a group can be demanding in practice, and the conditions under which a designer is invited to do this kind of work shape how well they can later take part in the group's thinking.

Our position in this paper is that AI in design ideation might be more usefully treated as a friction agent for reflection rather than as a smoothing agent for output. 
By introducing structured pauses that invite designers to articulate, monitor or commit, AI could support the reflection-in-action through which \textit{rationale} is built, instead of bypassing it. 
We outline what this reframing could involve, drawing on previous work on metacognition, group cognition and our own work on human-AI interaction, and surface the questions it opens up for the workshop community.

\section{Rationale as the Currency of Design Ideation}
Design ideation is commonly described as a movement between divergent and convergent thinking, where designers generate many candidates before narrowing toward the ones worth pursuing~\cite{gonccalves2016inspiration}. 
This framing focuses attention on the ideas themselves and less on what designers build alongside them. 
Through the act of accepting, rejecting and reworking candidates, designers also develop a rationale, the reasoning that explains why a candidate is worth keeping, dropping, combining or pushing on~\cite{maclean2020questions}. 
This rationale belongs to what ideation produces. 
It allows a designer to return to an idea later, to recognise when a seemingly new direction recombines older ones, and to know which constraints are load-bearing and which remain open to negotiation.

In individual ideation, weak rationale carries mostly internal costs. 
A designer who cannot reconstruct why they made a choice may struggle to revise it deliberately and may retrace the same exploration without learning from it. 
These costs are real yet often hidden, and they become structural the moment the designer enters a group. 
Group ideation depends on a continually constructed shared conception of the problem~\cite{roschelle1995construction, koch2018group}, and this conception rests as much on the accounts designers give of their contributions as on the contributions themselves. 

Giving such accounts involves two kinds of work.
Designers first build the rationale, connecting an idea to others that have been considered, weighing its strengths and limitations against similar but rejected directions, and locating it in relation to the problem it addresses. 
They then express it in a form collaborators can engage with, choosing a suitable level of detail and the references or framings that convey the reasoning rather than only the surface idea. Both kinds of work are demanding across levels of experience. 
Each project brings its own internal references, norms and ways of seeing, and collaborators arrive with their own perspectives, prior ideas and background knowledge. 
This diversity often fuels productive discussion, yet it also risks team members talking past each other when rationale stays thin or implicit.



Both kinds of work also come with their own challenges when ideas move into a group setting.
A contribution whose rationale cannot be articulated is difficult to defend, contest, combine or evolve with others, which means that a designer entering group ideation without rationale enters it without much of the means to participate fully. 
The materials designers bring with them carry their own limitations, since artifacts such as post-it notes, sketches or single images are by design reductive carriers~\cite{inie2017typology}. 
They are easy to produce, move and rearrange, and by the time an idea reaches the wall, most of the reasoning that produced it has been stripped away. A post-it shows what the idea is, not why it is there.

From a distributed cognition perspective, the cognitive work of making sense of ideas is carried by people and artifacts together rather than by individual minds alone~\cite{Hollan2000Distributed}. 
In design ideation, rationale spreads across sketches, written notes, collaged materials and digital prompts, which designers continuously merge and reconfigure based on perceived relationships between ideas. 
A single post-it or image shows only the current surface of an idea, and what a team ``knows'' about why it exists lies embedded in these distributed traces of activity. 
Designers often supply the missing context in conversation, so that the artifact serves as a handle while the designer carries the rest. 
When rationale is weak, the artifact has to stand alone, and ideas may be adopted, dropped or merged based on how well they read on a wall rather than on the reasoning behind them. 
Supporting design ideation therefore involves more than helping designers produce ideas. 
It also means helping them build rationale they can reflect on and share, in forms that others can engage with as more than surface artifacts.

\section{The Friction-Removal Trap of AI}
Yet most current generative AI tools for creative work tend to be designed around an implicit goal of throughput. 
Prompts are answered quickly, candidates are generated in bulk, and rejection is made about as cheap as acceptance. In many ways this is a feature, since designers can produce many candidates in a short time. 
From the perspective of rationale, however, it can become a problem, since designers may move through long sequences of generation and selection without having to articulate why one direction matters more than another. 
For example, designers frequently break existing artifacts into meaningful units, select fragments from different versions, and collage or assemble these into new configurations. 
Through this ongoing work of breaking, selecting and recombining, they actively shape and refine their design rationale, deciding what to keep, what to discard, and how different directions relate to one another. 
Current AI ideation tools, however, focus on the final steps, converting semantic prompts into visual output, and offer less help with the reflective work of segmenting and selecting elements in the moment. 
As recent position and empirical papers on AI‑supported ideation and UI design workflows argue, most generative tools are optimised for rapid generation and transformation at the level of completed artifacts, while offering little support for the fine‑grained sense‑making through which designers organise fragments, compare directions, and build rationale~\cite{wadinambiarachchi2026reviving,Petridis2023-wn}.
Some of our own work points in a similar direction, where designers given richer means of expression, such as multimodal prompting~\cite{peng2024designprompt} or pen-based composition-as-prompt~\cite{peng2025fusain}, often prefer interactions that are more demanding rather than less, which suggests that fluency on its own may not be the goal designers themselves are optimising for.

The risk, then, is not only that AI takes over labour that designers used to do. 
It is also that it reduces the reflection-in-action through which rationale is normally built alongside the artifact. 
With less of this in-the-moment reflection, the resulting rationale tends to be thinner and narrower in scope, and recent empirical work suggests that even basic recall of one's own contributions can suffer in mixed human-AI workflows~\cite{Zindulka2026}. 
The consequence is less that ideas become worse and more that potentially interesting ideas may struggle to find their place in the wider ideation narrative of the problem.
Without the connective reasoning that ties an idea to what came before and what it is responding to, a contribution can feel disconnected from the shared conception of the problem the group is constructing, and may be set aside or merged on the basis of how it presents rather than what it is doing in the design space.

\section{Friction as Scaffolding for Metacognition}
To support reflection in design ideation more meaningfully, it helps to be precise about which kind of reflection we consider.
Schön's distinction between reflection-in-action and reflection-on-action is useful here~\cite{schon2017reflective}. 
Reflection-in-action describes the live monitoring and adjustment of one's own thinking during an activity, where a designer notices something unexpected and reorients in the moment. 
Reflection-on-action describes the retrospective sense-making that happens afterwards, when a designer looks back at what they did and tries to understand it more systematically. 
Both are forms of metacognitive engagement, in the broader sense of monitoring and exerting control over one's own cognitive processes~\cite{flavell1979metacognition, nelson1990metamemory}, and both produce different kinds of rationale, the in-action kind being situated and tied to specific moments of choice, the on-action kind being more synthesised and easier to communicate later~\cite{schon2017reflective}. 

We argue that reflection-in-action is where the rationale of design ideation is most directly built, and also where current AI tools tend to be least supportive. 
Existing reflective creativity support tools have largely focused on reflection-on-action, by helping designers document, revisit and visualise their process after the fact~\cite{Sharmin2013ReflectionSpace,Sterman2023Kaleidoscope}. 
These tools are valuable, but they often leave the in-the-moment work rather unsupported, and in some cases encourage designers to defer reflection to a later stage. 
Some recent work has begun to investigate how AI itself could support reflection in the moment in human-AI co-creation~\cite{Wang2025Pinning,Xu2025Productive}, though this remains a less developed area.
AI sits inside the creative loop, where the in-action reflection happens, which makes it well positioned to support that reflection rather than to bypass it, provided that the interaction is designed to invite articulation rather than to resolve it on the designer's behalf.\\

We propose AI as friction as a way to address this.
With friction we mean a structured pause that invites the designer to articulate, monitor or commit (aligned with~\cite{Cox2016Design}), rather than an obstruction that simply slows the interaction down. 
Some illustrative mechanisms might give a sense of what such friction could look like in practice: 
An AI agent might ask a designer to briefly articulate why an idea is being rejected before it leaves the canvas, so that the act of dismissal becomes a moment of reasoning. 
It might also surface a contrast or a counter-example rather than another continuation, so that the designer has to position their current direction against an alternative. 
It could also hold back a generation until the designer commits to a direction, so that a prompt becomes a small act of intention rather than a reflex. 
Or it might reflect back patterns in what is being accepted and rejected over time, so that the designer can become aware of their own implicit criteria. 
A further variant is an agent that behaves counterintuitively compared to conventional one-click generation tools. 
Instead of turning a prompt or sketch into a single finished image, it returns a set of visual units that designers must tweak, move and recombine before any complete artifact appears. 
The extra work with these intermediate fragments itself becomes friction that keeps designers in the reflective process of breaking, selecting and merging where new ideas and rationale emerge.
These are sketches, not proposals, highlighting the larger idea that AI is uniquely positioned to introduce such moments within the loop, where reflection-in-action would otherwise be easily skipped.

We want to be clear that we understand friction as compatible with flow. It is a structured pause that converts process into rationale, and the boundary between productive friction and disruptive obstruction needs to be designed and studied carefully~\cite{chen2024exploring}. 
The in-action reflection that friction supports is also what later forms of reflection rest on. 
Rationale that is built moment by moment is what reflection-on-action can later draw upon, and what allows the resulting artifact to travel into group ideation together with the reasoning that produced it.

\section{Open Questions for the Workshop}
Several questions remain open in this framing, and we would like to bring them into the workshop discussion. 
The first addresses whether and how friction can be introduced in practice without disrupting the very process it is meant to enrich.
\textit{How much friction is productive, and how would we know?}
\begin{itemize}
    \item Where does the boundary lie between a structured pause that invites articulation or reflection and an obstruction that slows the designer down or causes frustration?
    \item Is the trade-off between creative flow and reflection as sharp as it is sometimes assumed, or does productive friction work with rather than against flow under the right conditions?
    \item How might friction-for-reflection be evaluated, given that its benefits tend to accrue over time and across the solo-to-group transition rather than within a single session, where session-level metrics such as task completion or output quantity may not be relevant?
\end{itemize}

\noindent A second set of questions concerns the shape that friction should take when ideation moves from individual work into group settings, where rationale has to travel as well as be built and developed further.
\textit{What should friction look like in groups, and for whom?}
\begin{itemize}
    \item Could AI play a role in helping designers carry their rationale forward into group ideation, perhaps by acting as a kind of translator between an individual's reasoning and the shared narrative the group is constructing?
    \item Group ideation involves different participants moving in and out of leading and following roles, so uniform friction across collaborators is unlikely to be appropriate. What might adaptive friction look like, and how could an AI agent decide who to direct it toward, and when?
    \item How might friction support the group's shared conception of the problem itself, rather than only individual rationale, for instance by surfacing divergence between collaborators' implicit criteria?
\end{itemize}

\noindent These questions reflect what we see as the most interesting design space opened up by treating AI as a friction agent for reflection rather than a smoothing agent for output, and we look forward to thinking through them with the workshop community.

\begin{acks}
This work was supported by a French government grant managed by the Agence Nationale de la Recherche as part of the France 2030 program, reference ANR-22-EXEN-0004 (PEPR eNSEMBLE / PC3 MATCHING)
\end{acks}

\bibliographystyle{ACM-Reference-Format}
\bibliography{sample-base}


\end{document}